# Comparing autonomous vehicle acceptance of German residents with and without visual impairments

Kacperski, Celina[1,2] Kutzner, Florian[1] Vogel, Tobias[3]

[1] Seeburg Castle University

[2]University of Konstanz, Cluster of Excellence

[3] University of Applied Sciences Darmstadt

e-mail: celina.kacperski@uni-seeburg.at

**Acknowledgements**

Financial support by the European Union Horizon 2020 research and innovation programme is gratefully acknowledged (Project PAsCAL, grant N°815098).

We would like to thank the DBSV (Deutscher Blinden- und Sehbehindertenverband) and the European Blind Union (EBU) for their support in data collection, and our colleagues at EBU for their helpful comments on a draft of this paper.

Abstract

Connected and autonomous vehicles (CAVs) will greatly impact the lives of individuals with visual impairments, but how they differ in expectations compared to sighted individuals is not clear. The present research reports results based on survey responses from 114 visually impaired participants and 117 panel recruited participants without visual impairments, from Germany. Their attitudes towards autonomous vehicles and their expectations for consequences of wide-spread adoption of CAVs are assessed. Results indicate significantly more positive CAV attitudes in participants with visual impairments compared to those without visual impairments. Mediation analyses indicate that visually impaired individuals' more positive CAV attitudes (compared to sighted individuals') are largely explained by higher hopes for independence, and more optimistic expectations regarding safety and sustainability. Policy makers should ensure accessibility without sacrificing goals for higher safety and lower ecological impact to make CAVs an acceptable inclusive mobility solution.

# Comparing autonomous vehicle acceptance of German residents with and without visual impairments

Many people with disabilities face limited transportation options (Brumbaugh, 2018; Dicianno et al., 2021; Riggs & Pande, 2022). Among them, individuals with visual impairments are uniquely affected, as they often do not own a car, and if they do, driving at night, in rain, or in heavy traffic can be challenging (Bezyak et al., 2017). At the same time, sight loss is not a rare occurrence: over 10% of Europeans experiences some form of visual impairment in their lifetime (World Health Organization, 2019). Individuals with visual impairments are often reliant on assistance for their transportation needs; they can be shut out from employment and leisure activities, experience social isolation and an overall lower quality of life (Claypool et al., 2017; Crewe et al., 2011; Crudden et al., 2015; Kempen et al., 2012; Montarzino et al., 2007). Frustration due to the dependence on others, as well as fear of unfamiliar environments are commonly reported emotions (Crudden et al., 2015; Montarzino et al., 2007).

A variety of potential solutions are discussed to increase the autonomy of individuals with disabilities and create accessible forms of transportation. One example is a transport system with increased geographic, temporal or vehicle flexibility, including tailored services for vulnerable users (Cottrill et al., 2020). Technological solutions have been proposed, mostly in the context of smart city development, where better navigation services could be integrated via the Internet-of-Things (Gong et al., 2017). Connected and autonomous vehicles (CAVs) are part of this flexible technological development. Integrated into larger existing mobility infrastructure, they present a unique opportunity to increase life quality and transport experience of visually impaired citizens.

Taking into account the assumed immensity of the impact of CAVs on the lives of people with visual impairments (Brinkley et al., 2017; Fink et al., 2021), previous studies have looked into their attitudes and expectations (Bennett et al., 2020; Brinkley et al., 2020; Fink, Alsamsam, et al., 2023; Miller et al., 2022), and recently, research has begun to proliferate that focuses on human machine interaction (HMI) design issues (Fink, Dimitrov, et al., 2023). Here, the aim is to improve assistive technology, and accessible onboard and autonomous systems, through multisensory interfaces (Fink, Dimitrov, et al., 2023), haptic designs and reduction of visual cues (Brewer & Kameswaran, 2018; Brito et al., 2018; Ranjbar et al., 2022; Sucu & Folmer, 2014) and vehicle assistance, including voice controlled facilities, sensor fusion approaches, or Braille inspired tactile interfaces (Brito et al., 2018; Fink, Doore, et al., 2023; Hong et al., 2008; Sucu & Folmer, 2014). However, to the best of our knowledge, no research so far exists investigating differences between visually impaired and sighted individuals' acceptance of CAVs, and how these differences are motivated. Thus, we will first give an overview over literature showcasing determinants of CAV acceptance of individuals with and without visual impairments, and then present the proposed research.

## Autonomous vehicle acceptance of individuals with and without visual impairments

Qualitative studies have provided exploratory insights into the needs and expectations of individuals with visual impairments. Here, themes of independence leading to a better quality of life are frequent (Brewer & Kameswaran, 2018). An optimistic perspective abounds, in particular regarding reduction of mobility barriers and increase of autonomy and self-determination (Brinkley et al., 2017). Concerns have been reported regarding control and safety of the developed technology and that it might not satisfy needs for situational awareness and mobility assistance (Brewer & Kameswaran, 2018; Brinkley et al., 2017; Miller et al., 2022).

However, in vivo experience with CAV prototypes have shown an increase in trust, reduced fear, belief in usability and in the desire to purchase (Brinkley, Posadas, et al., 2019; Fink, Dimitrov, et al., 2023). These are effects similar to those observed in non-visually impaired individuals after a CAV experience (Eden et al., 2017; Liu, Xu, et al., 2019; Liu & Xu, 2020). When a quasi-experimental design study compared facial expressions and self-reported emotions of individuals with and without visual impairment when taking part in a CAV test ride, no significant differences were reported: over 90% of registered facial expressions indicated happiness or surprise, and participants of both groups reported highly enjoyable experiences and feelings of safety, with very little anxiety and fear (Kempapidis et al., 2020).

Four quantitative studies resonate with the qualitative findings. Descriptively, individuals with visual impairments reported relatively high hopes regarding accident frequency, environmental benefits such as lower vehicle emissions and better fuel economy, and vehicle efficiency and reliability (Miller et al., 2022). And, as before, these were paired with safety concerns regarding equipment failure, cyber security and pedestrian safety (Brinkley & Gilbert, 2018) as well as control and communication issues (Miller et al., 2022). Open answers from people who are blind, which were quantitized, indicated hope for more independence, safety improvements, and affordability predicting willingness to travel in CAVs (Bennett et al., 2020). And, a stated preference choice experiment found accessibility, safety and reliability as major predictors of CAV usage intentions, with CAVs more frequently chosen than other modes of transportation, indicating a strong willingness to use them (Hwang et al., 2020).

In sum, evidence on the attitudes and acceptance of CAVs by visually impaired people is still often exploratory, and studies with quantitative evidence, and systematic comparisons with samples of non-visually impaired persons are missing (Miller et al., 2022). Additionally, there is a paucity of experimental studies focused on the design and testing of interfaces from the perspective of users with visual impairments: Fink et al. (2021; 2023) identify this gap and point toward the need for more empirical research that can inform regulatory frameworks and guide the development of visual-impairment-centric CAV technologies.

Research in the general population is more comprehensive (for literature reviews, see Golbabaei et al., 2020; Jing et al., 2020; Keszey, 2020). Similar to studies with visually impaired individuals (Brewer & Kameswaran, 2018; Brinkley, Posadas, et al., 2019; Sucu & Folmer, 2014), research often concentrates on immediate user-experience factors such as ease of use and usefulness (Jing et al., 2020; Nordhoff et al., 2018, 2020; Sener et al., 2019), with trust in the vehicle a major indicator of willingness to use CAVs (Jing et al., 2020; Liu, Yang, et al., 2019; Nordhoff et al., 2020; Zhang et al., 2019). However, recent research based on the Theory of Planned Behavior (TPB, Ajzen, 1991) and the Unified Theory of Acceptance and Use of Technology (see TAM, UTAUT2, Davis, 1989; Venkatesh et al., 2012) has broadened the scope and assessed a wider range of aspects underlying CAV acceptance (Kacperski et al., 2021).

Specifically, an effort was made to assess various expected consequences, which, based on the TPB, can be considered central to attitudes (Ajzen, 2002; Vogel & Wänke, 2016). Four factors of expected consequences emerged: privacy, safety, sustainability, and efficiency. All four factors predicted CAV evaluations to some degree - safety and efficiency predicted general evaluations, while affective evaluations were predicted by all four factors, with safety and privacy having the strongest effects. While consequences of CAV introduction had previously been studied in term of concerns about travel safety (Jing et al., 2020; Montoro et al., 2019), consequences for the environment (Jing et al., 2020; Wu et al., 2019) and economic and time savings (Jing et al., 2020; König & Neumayr, 2017), a joined investigation of a wide range of potential consequences allowed for a better understanding of their relative importance.

**Proposed research**

The present research aims to build on previous research conducted on attitudes by individuals with visual impairments (Bennett et al., 2020; Miller et al., 2022) and extend on it by integrating and comparing the perspectives of both visually impaired and sighted persons with quantitative evidence. First studies have identified safety, independence and a positive effect on the environment (pollution) as potential expected benefits from individuals with visual impairments (Brewer & Kameswaran, 2018; Brinkley et al., 2017, 2020), and we aim to contribute here by providing a structured and in-depth overview of expectations of individuals with visual impairments with CAVs.

For RQ1, we expect to replicate the four factors (privacy, safety, sustainability, and efficiency) found previously (Kacperski et al., 2021) and to see a good fit when including independence as an additional factor, due to its high relevance for visually impaired people. We will employ Confirmatory Factor Analysis (CFA) to verify factor structure.

For RQ2, we will assess differences in expectations held by persons with visual impairments as compared to those held by representative panel participants, through descriptive analyses. As part of this research question, we propose the following hypotheses:

H1: We hypothesize, based on previously reported optimistic mind-sets of individuals with visual impairments towards CAVs (Brewer & Kameswaran, 2018; Brinkley et al., 2017; Brinkley & Gilbert, 2018), that we will find them to have an overall more positive evaluation of CAVs, as compared to the panel sample.

H2: We also expect that independence benefits will play are larger role for visually impaired participants. On the other factors, differences are investigated in an exploratory manner.

We will also explore polarization among our participants with regards to the five identified factors (independence, privacy, safety, sustainability, and efficiency).

For RQ3, we will explore which consequence factors better explain more positive evaluations for individuals with and without visual impairments. For this, we employ mediation models.

# Methods

## Participants

We collected data from two sources, both in Germany. The DBSV (Deutscher Blinden- und Sehbehindertenverband), supporting the European Blind Union, recruited 156 German participants with visual impairments via their online social networks without compensation by inviting them to the survey via a link. For the following, we will use the definition by the WHO (2019) of visual impairment, which "occurs when an eye condition affects the visual system and one or more of its vision functions".

Simultaneously, a German panel provider collected 120 German participants via the same link; compensation was offered in line with panel provider guidelines.

We defined two exclusion criteria: for the sample of participants with visual impairments, we excluded all participants who reported not having a visual impairment (30). For both samples, we excluded participants not between the ages of 18 and 67 (15), i.e., what is generally considered the "working age population". While most individuals with visual impairments are over 60 years of age, this decision was made due to difficulty of obtaining data from participants in higher age groups. Generalizations of results beyond a "working age population" should therefore be made with care.

We retained n=114 participants (50 women) with visual impairments, and n=117 participants (51 women) from the panel sample. Demographics can be found in Table 1. Participants with visual impairments were on average slightly older, were more likely to hold educational level of high school and above, and less likely to have net household income in the lowest (below 1000 Euro) or highest bracket (above 3000 Euro). They were also far less likely to hold a driver's license or own a car than participants from the panel sample.

*Table 1. Sociodemographic and mobility data.*

| Variables | Participants with visual impairments | | | Panel participants | | |
|---|---|---|---|---|---|---|
| **Age Group** | **18-29** | **30-49** | **50-67** | **18-29** | **30-49** | **50-67** |
| | 9 | 51 | 54 | 24 | 53 | 40 |
| **Gender** | **Women** | **Men** | | **Women** | **Men** | |
| | 50 | 64 | | 51 | 66 | |
| **Education** | **Low** | **Middle** | **High** | **Low** | **Middle** | **High** |
| | 14 | 28 | 72 | 17 | 55 | 45 |
| **Income**[1] | **< 1000** | **1000-3000** | **> 3000** | **< 1000** | **1000-3000** | **> 3000** |
| | 11 | 63 | 18 | 21 | 54 | 32 |
| **Driver's License** | **No** | **Yes** | | **No** | **Yes** | |
| | 92 | 22 | | 6 | 111 | |
| **Car owned** | **No** | **Yes** | | **No** | **Yes** | |
| | 83 | 31 | | 24 | 93 | |

*Note: Education (low = elementary school and below; high = high school and above; middle = middle school, apprenticeship, vocational school); income in Euros.*

[1] 33 participants (22 visually impaired) chose not to provide information on this question

Ages ranged from 18 to 67 with M = 47.5 (SD =12.3) for participants with visual impairments, and M = 43.0 (SD = 13.9) for participants from the panel sample. 0.01% of participants reported being deaf blind (n=1), 65.87% reported being blind (n=75), and 33.33% reported being partially sighted (n=38). None of the participants had experience with autonomous vehicles.

## Survey

### Procedure

Ethics requirements of the German Ethics Board (DGPS) as well as European data protection guidelines (DGPR) were observed. Participants were invited via a link to the programmed survey on SoSciSurvey, a Germany-based host provider. Image descriptions and questionnaire items were provided in such as a manner that the survey was accessible with screen readers. The survey on personal and large-scale consequences of CAV proliferation started with an introductory page with a brief welcome and study purpose, as well as a letter of information and consent, including information that participants could leave the survey at any time.

Then, participants received a short vignette informing about the functioning of a self-driving car (Level 5)[2] together with an icon, and the description of the presented icon ("a sedan with a wireless signal emitting from it"). On the following pages, we presented participants with items for general and affective evaluations regarding CAVs, items on personal and large-scale consequences, items surveying intention to use CAVs, mobility behaviours, and socio-demographics. For more details, please see Kacperski et al., 2021.

[2] "In the following we will ask you some questions about autonomous and connected vehicles (Connected Autonomous Vehicle, CAV for short). The distinctive feature of a CAV is that it is not controlled by a human driver. Instead, it is completely controlled by a computer system. The vehicle takes over all driving tasks and automatically controls all actions, including steering, acceleration and braking."

Participants on average spent ca. 40 minutes on the survey, with panel participants averaging 20.87 minutes (SD = 7.08) and participants with visual impairments averaging 48.81 minutes (SD = 22.71).

**Survey items**

Items on CAV consequences were adapted from literature on predicted consequences of CAV propagation (Anderson et al., 2016; Kassens-Noor et al., 2020; Narayanan et al., 2020; Nordhoff et al., 2016) and based on insights from qualitative interviews carried out independently (Kacperski et al., 2020). Each consequence item was comprised of the evaluation, and its importance for the participant (Krosnick & Abelson, 1992; Vogel & Wänke, 2016), using semantic differentials with seven-point scales. For example, with reference to the environmental consequences, "If large sections of the population used connected and autonomous vehicles, the environment would be doing [worse to better]." To measure importance, we asked to rate the statement "The fact that the environment is doing well is [unimportant to important] to me." Another example related to social consequences is, "If I used an autonomous car, meetings with acquaintances (e.g., friends, family) ... would be [less frequent to more frequent].", and for importance, we asked "Frequent meetings with acquaintances are [unimportant/important] to me." For general evaluation, an example would be, on a seven-point scale, "In principle, I find connected and autonomous cars [very bad to very good]." An example for the affective evaluation seven-point scale is, "The idea of large sections of the population using connected and autonomous cars feels good. [disagree completely to agree completely]". We also asked for prior experience with autonomous cars. Finally, visual impairment was assessed based on self-report, with the options "I am blind", "I am partially-sighted" and "I am deaf-blind". Appendix A holds the full list of items.

**Data analysis**

Survey responses were examined for data quality and exclusions. Data was handled with R statistics (R Development Core Team, 2008). Consequence items coded from -3 to 3 were weighted with perceived importance ratings from 1 to 7, creating a measure weighting the expected consequences (Ajzen & Fishbein, 2008; Krosnick & Abelson, 1992); non-weighted raw values were used for descriptive statistics.

First, we performed a confirmatory factor analysis to evaluate the measurement relationships between the latent and observed variables (Hair et al., 2018). We report psychometric properties such as reliability (Cronbach's alphas > 0.70), convergent validity (scale loadings > 0.50) and discriminant validity (heterotrait-monotrait criteria (HTMT) < 1) and appropriate fit indices (Hair et al., 2018).

Following the CFA, descriptive statistics such as means and standard deviations were calculated. Group differences between participants with visual impairments and the sample were conducted with analyses of variance (using lme/aov in R), controlling for age, gender, income and education levels. Finally, a mediation was modelled with lavaan (Rosseel, 2012), with assesses the significance of pathways by bootstrapping, to evaluate the relative impact of the found factors on CAV acceptance, depending on sample, controlling for age, gender, income and education levels.[3]

[3] Full output of the mediation models can be found in Appendix B.

# Results

## Confirmatory factor analysis

The original 53 item pairs were reduced to 19[4], with five confirmed factors as can be seen in Table 2. The item "speed of vehicle" was excluded from the factor Efficiency and from further analysis due to a factor loading below 0.30. Of the original 12 items available for the Independence factor, 6 were selected due to highest factor loadings and for yielding the best model fit overall; they represent impact of CAV introduction on job opportunities, socializing opportunities and freedom of decision-making. All items in the final model loaded > 0.5 on their factors.

*Table 2. Latent factors with items, their loadings and descriptives.*

| Latent factor | Item | Loading | Mean | SD |
|---|---|---|---|---|
| Privacy | Data safety | 0.84 | -1.20 | 1.62 |
| | Data abuse | 0.83 | -1.28 | 1.63 |
| | Surveillance | 0.76 | -1.47 | 1.66 |
| Sustainability | Environmental cost | 0.87 | 0.84 | 1.68 |
| | Environmental degradation | 0.86 | 0.87 | 1.62 |
| | Emissions | 0.89 | 0.76 | 1.66 |
| | Pollution | 0.92 | 1.00 | 1.66 |
| Safety | Road safety | 0.82 | 0.49 | 1.99 |
| | Accident risk | 0.76 | 0.46 | 1.93 |
| | Number of accidents | 0.86 | 0.93 | 1.83 |
| | Danger of travel | 0.88 | 0.69 | 1.82 |
| Efficiency | Trip duration | 0.88 | 0.60 | 1.68 |
| | Travel speed | 0.87 | 0.56 | 1.84 |
| | Travel time | 0.75 | 0.51 | 1.70 |
| Independence | Freedom of decision | 0.79 | 0.45 | 2.01 |
| | Socializing at events | 0.83 | 0.62 | 1.66 |
| | Socializing with peers | 0.86 | 0.87 | 1.61 |
| | Job availability | 0.75 | 0.45 | 1.43 |
| | Job opportunity | 0.79 | 0.60 | 1.62 |
| | Job productivity | 0.66 | 0.42 | 1.59 |

[4] The list of selected items can be found in Appendix A.2

Note: Means based on raw values (range: -3 to 3). Items asked for expected consequences of CAV usage, with positive (negative) values indicating an aspect would improve (worsen).

Acceptable discriminant validity was determined for this solution (Hair et al., 2018), with heterotrait-monotrait (HTMT) criteria for each pair of constructs on the basis of item correlations values between 0.37 (for Privacy/Sustainability) and 0.88 (for Efficiency/Independence), clearly smaller than 1 (Henseler et al., 2015), see Figure 3.

*Table 3. Factor correlations and AVEs.*

| Construct | Privacy | Sustainability | Safety | Efficiency | Independence |
|---|---|---|---|---|---|
| Privacy | *0.66* | 0.37 | 0.41 | 0.47 | 0.46 |
| Sustainability | | *0.78* | 0.51 | 0.47 | 0.45 |
| Safety | | | *0.69* | 0.73 | 0.69 |
| Efficiency | | | | *0.70* | 0.87 |
| Independence | | | | | *0.61* |

*Note*: Diagonals are average variance extracted (AVEs). The off-diagonal elements are the correlations between the latent factors.

Fit measures indicate an adequate to good fit as reported in Table 4 alongside commonly acceptable cutoff values (Hooper et al., 2007; Kline, 2005). Reliability was also considered acceptable with Cronbach's alphas of all constructs above 0.79 (see Table 5).

*Table 4. Fit indices for CFA and lavaan path model (SEM).*

| Measure | CFA | Cutoff |
|---|---|---|
| CFI | 0.93 | > 0.9 |
| TLI | 0.92 | > 0.9 |
| RMSEA | 0.08 | < 0.08 |
| SRMR | 0.06 | < 0.08 |
| Relative $\chi^2$ | 2.46:1 | < 3:1 |
| *($\chi^2/df$)* | *393/160* | |

**Descriptive analyses**

Unlike for panel participants, visually impaired participants' ratings for both general and affective evaluation of CAVs, consistently stayed above the scale midpoints (see Table 5). Additionally, expected consequences of CAVs were rated positively by visually impaired participants except for consequences related to privacy; panel participants were neutral regarding efficiency and independence, i.e. they expected neither improvement nor worsening.

*Table 5. Raw score means and standard deviations for consequence and importance items.*

| Scales (midpoint) | | α | Mean | SD | SE | Mean | SD | SE |
|---|---|---|---|---|---|---|---|---|
| | | | Participants with visual impairments | | | Panel participants | | |
| Evaluation (4) | General | 0.96 | 5.36 | 1.82 | 0.17 | 3.86 | 1.87 | 0.17 |
| Evaluation (4) | Affective | 0.94 | 5.11 | 1.55 | 0.15 | 3.87 | 1.77 | 0.16 |
| Consequence (0) | Privacy | 0.84 | -1.11 | 1.46 | 0.14 | -1.52 | 1.37 | 0.13 |
| Consequence (0) | Sustainability | 0.93 | 1.13 | 1.55 | 0.15 | 0.61 | 1.41 | 0.13 |
| Consequence (0) | Safety | 0.90 | 0.92 | 1.65 | 0.15 | 0.37 | 1.64 | 0.15 |
| Consequence (0) | Efficiency | 0.85 | 1.13 | 1.38 | 0.13 | -0.01 | 1.46 | 0.14 |
| Consequence (0) | Independence | 0.89 | 1.09 | 1.33 | 0.12 | 0.06 | 1.14 | 0.11 |

*Note. Alpha (α) is the Cronbach's alpha for each factor. SD is the standard deviation. SE is the standard error. Evaluation ratings ranged from 1-7. Consequence valence ratings ranged from -3 to 3; Scale midpoints are indicated in brackets behind scale name.*

Figure 1 illustrates raw mean scores across the main consequence scales, comparing means between participants with visual impairments and from the panel.

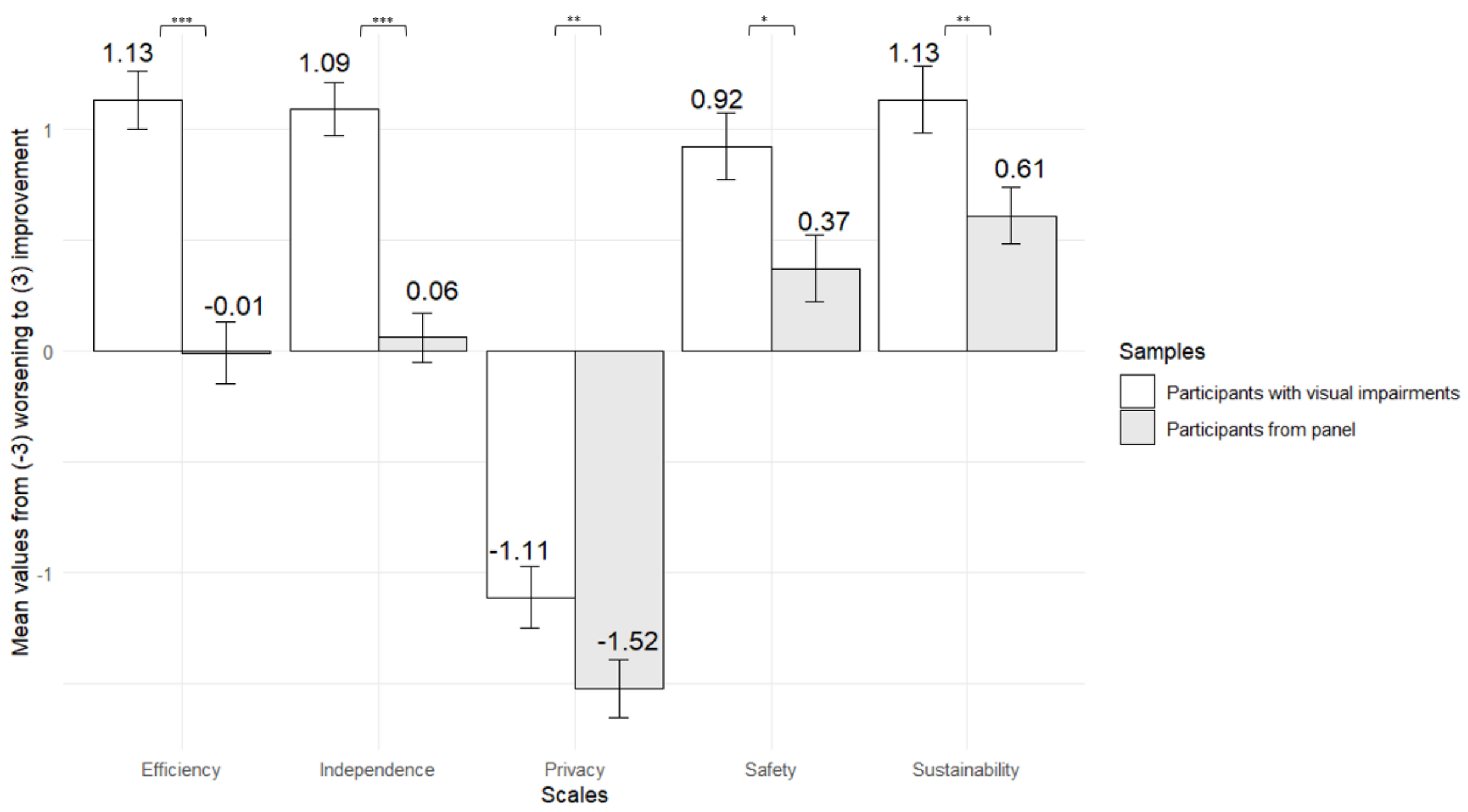

*Figure 1. Average consequence expectations for five factors – Efficiency (i.e. higher = expected improvement of vehicle speed and travel time); Independence (i.e. higher = expected improvement of economic and social participation); Privacy (i.e. higher = expected improvement of data protection and surveillance); Safety (i.e. higher = expected improvement of road safety); Sustainability (i.e. higher = expected improvement of emissions or pollution). Each scale shows the values for the two samples (participants with visual impairments (white) and participants from panel (grey)) and significance brackets with p-values (*** < 0.001; ** < 0.01; * < 0.05. Error bars represent standard errors. Table 5 summarizes displayed means, standard deviations, and standard errors.*

Table 6 shows results of an analyses of variance on aggregate scores of the items of the displayed factors, comparing the two samples when controlling for age, gender, education, and level of income.

*Table 6. ANCOVAs for differences between participants with visual impairments and the sample of panel participants; raw scores were used for all variables; all tests controlled for age, gender, education and income. No significant effects of education and income were found.*

| Scales (midpoint) | | SS | Sample F | Sample $\omega^2p$ | Age F | Gender F |
|---|---|---|---|---|---|---|
| Evaluation (4) | General | 103.771 | 33.047*** | 0.122 | 6.623* | 11.787*** |
| Evaluation (4) | Affective | 71.228 | 27.272*** | 0.102 | 5.666* | 11.942*** |
| Consequence (0) | Privacy | 14.956 | 7.570** | 0.028 | 5.293* | 0.365 |
| Consequence (0) | Sustainability | 17.999 | 8.169** | 0.030 | 0.507 | 1.383 |
| Consequence (0) | Safety | 10.426 | 4.098* | 0.013 | 3.400* | 14.162*** |
| Consequence (0) | Efficiency | 58.789 | 31.152*** | 0.115 | 12.378** | 3.260 |
| Consequence (0) | Independence | 50.916 | 35.648*** | 0.130 | 13.648*** | 1.713 |

*Note. Evaluations ranged from 1-7. Consequence valence ratings ranged from -3 to 3. Scale midpoints are indicated in brackets behind scale name. SS is sum of squares. Df (1,220). F-values include*

*significance levels (*** < 0.001; ** < 0.01; * < 0.05), partial omega squared ($\omega^2 p$) indicates effect size (> 0.01 small, >0.06 medium, > 0.14 large). Age (18-67, continuous) and gender (w(0), m(1)) are reported as F-values including significance levels. Detailed statistics for all variables can be found in Appendix C.1.*

We found that participants with visual impairments evaluated CAVs more positively, had a significantly higher intention to use CAVs, and expected CAVs to have a more positive impact on safety, privacy, sustainability, and efficiency as well as independence. We found the largest effect on the independence factor[5].

This general impression was confirmed by the differences in polarization regarding expected consequences, i.e., whether there was disagreement between participants regarding improvement or worsening in the various areas affected by CAV introduction. Figure 2 shows a bar plot of the polarization, with percentages of respondents that thought CAVs would improve (top bars) or worsen (bottom bars) the given area of impact (safety, sustainability, efficiency, privacy, independence).

There was very little polarization for participants with visual impairments. Across all five factors, over 65% of participants expected that privacy would worsen with CAV introduction, while safety, efficiency, sustainability and independence would improve. Less than 15% of participants felt that independence, sustainability and efficiency would worsen with CAV introduction. Only for safety, 22% of participants disagreed with the majority, instead expecting safety to worsen. This pattern is very different from panel participants, as they were divided regarding independence and efficiency (with over 30% each expecting improvement and worsening), safety (with 53% expecting improvement, while 34% expected worsening) and sustainability (with 55% expecting improvement and 21% worsening). Only with regards to privacy, most panel participants expected it to worsen (77%).

[5] We did not find significant effects of education and income. We did find that older participants expected more positive consequences, and they evaluated CAVs more positively; the latter was also true for men (as compared to women).

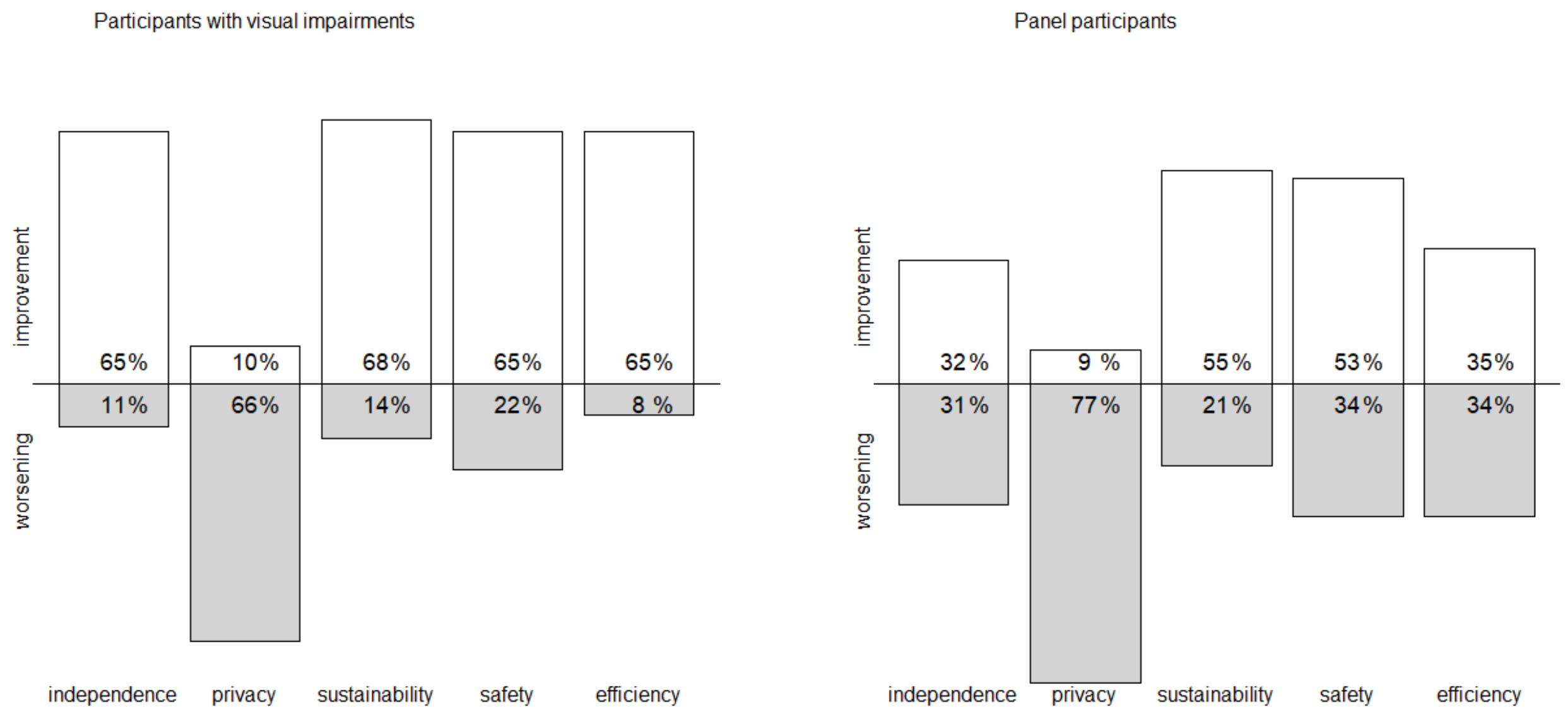

*Figure 2. Representation of polarization as a percentage of respondents expecting improvement (rating values > 0.5) or worsening (rating values < -0.5) from CAV introduction on the five factors: Independence, Privacy, Sustainability, Safety, Efficiency. Left hand graph shows participants with visual impairments, right hand graph shows panel participants.*

## Mediation analyses

Two parallel mediation models were conducted, for general and for affective evaluations. We tested to what extent the differences in evaluation of our two participant groups (visually impaired vs. panel participants) can be explained by differences in expectations about the five factors (efficiency, independence, privacy, safety, and sustainability). In both analyses, age, gender, education level and income level served as controls. As none of the control variables were significant, they will not be further discussed in the following.

Results (see Figure 3) indicate that independence and safety[6] were the two factors that mediated the relationship between sample and general evaluations, i.e., individuals with visual impairments evaluated CAVs more positively in general due to their expectation of higher independence and more safety. Additionally, we found that sustainability and safety mediated the relationship between sample and affective evaluation; individuals with visual impairments evaluated CAVs more positively in terms of affect, as they expected CAVs to increase sustainability and safety.

---

[6] The model shows significant a and b paths for privacy; however, the indirect effect was not significant (see Full model outputs of both mediation models can be found in Appendix C.2.

Table 7). This has been hypothesized to occur due to lack of shared variance for the two paths, or lack of power/inadequate sample size (Mackinnon, 2008).

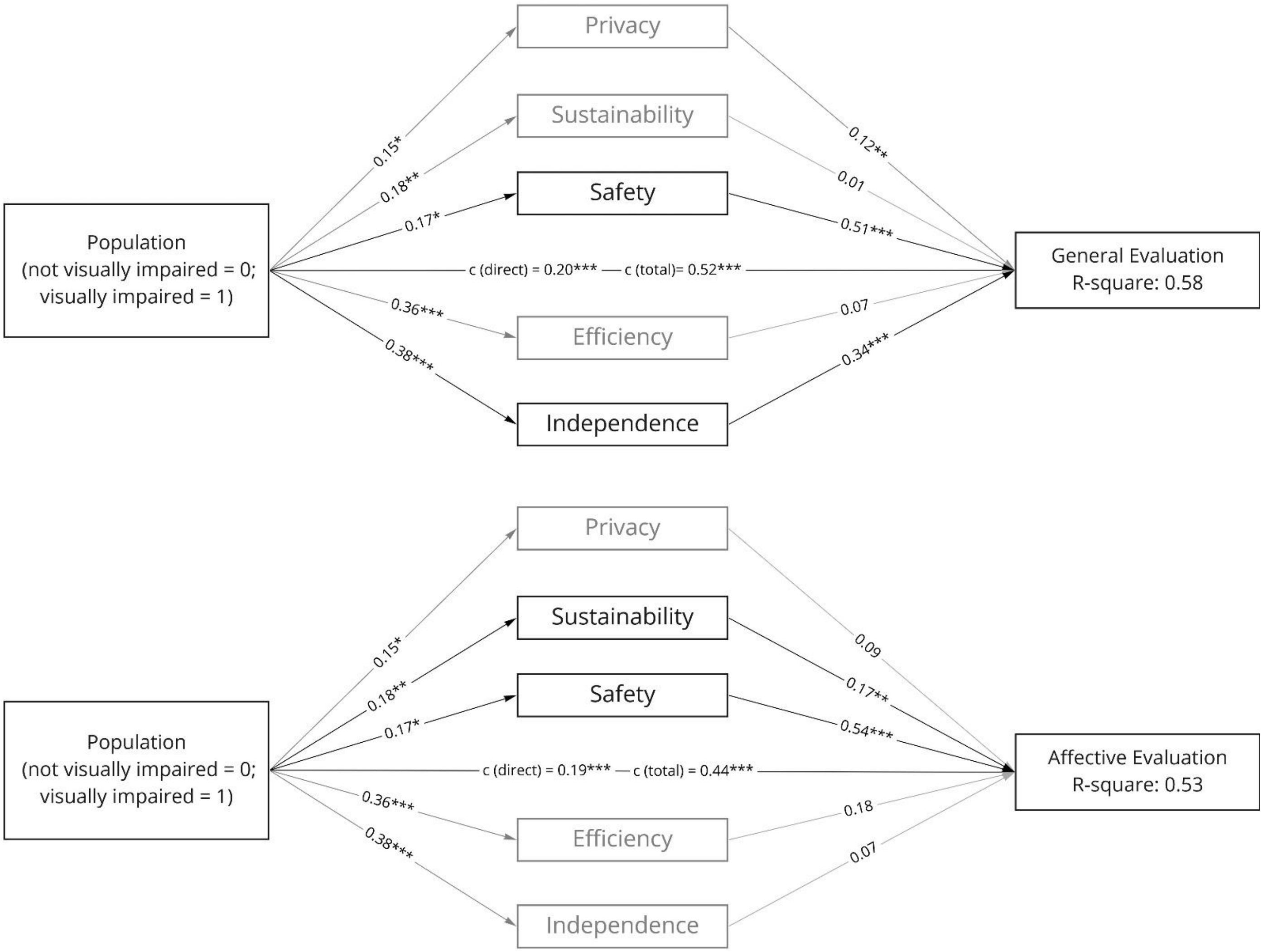

*Figure 3.* *Mediation analyses. The five factors mediating the relationship between population (participants with visual impairments and panel participants), and general as well as affective evaluations. Top graph: safety and independence are bolded due to their significant indirect effect on general evaluation. Bottom graph: sustainability and safety are bolded due to their significant indirect effect on affective evaluation.*

Full model outputs of both mediation models can be found in Appendix C.2.

Table 7 shows indirect effects for all five mediating factors. Full model outputs of both mediation models can be found in Appendix C.2.

*Table 7. Indirect effects of population (with visual impairment/without visual impairment) on privacy/sustainability/safety/efficiency/independence on general and affective evaluations.*

| | Factor | Std.all | SE | Wald z | p-value | 95% CI |
|---|---|---|---|---|---|---|
| General Evaluation | Privacy | 0.02 | 0.04 | 1.66 | 0.097 | [ 0.00, 0.15] |
| | Sustainability | 0.00 | 0.04 | 0.22 | 0.829 | [-0.07, 0.09] |
| | **Safety** | **0.08** | **0.12** | **2.30** | **0.021** | **[ 0.06, 0.55]** |
| | Efficiency | 0.02 | 0.11 | 0.73 | 0.468 | [-0.14, 0.32] |
| | **Independence** | **0.13** | **0.11** | **3.80** | **0.001** | **[ 0.22, 0.67]** |
| Affective Evaluation | Privacy | 0.01 | 0.03 | 1.30 | 0.193 | [-0.01, 0.11] |
| | **Sustainability** | **0.03** | **0.05** | **1.96** | **0.050** | **[ 0.01, 0.19]** |
| | **Safety** | **0.09** | **0.11** | **2.37** | **0.018** | **[ 0.06, 0.50]** |
| | Efficiency | 0.06 | 0.11 | 1.70 | 0.090 | [-0.02, 0.42] |
| | Independence | 0.03 | 0.11 | 0.78 | 0.438 | [-0.12, 0.30] |

*Note: Std.all are standardized beta coefficients. SE is the standard error. The p-value reports significances, with significant results bolded. The 95% CI is the confidence interval, bootstrapped 5000 times. Independence and safety show a significant indirect effect on general evaluation; sustainability and safety show a significant indirect effect on affective evaluation.*

Participants with visual impairments generally evaluated CAVs more positively; in addition, a more optimistic expectation for advances for their independence was the strongest mediator of this relationship, followed by higher expectations regarding safety. Similarly, participants with visual impairments generally had more positive affective evaluation of CAV; here, more optimistic expectation for safety was the strongest mediator, with sustainability also playing a role. We did not find a clear mediating relationship for privacy and efficiency, though participants with visual impairments had more optimistic expectations regarding consequences on these two factors.

## Discussion

Individuals with visual impairments are limited in their access to transportation options; connected and autonomous vehicles have been put forth as a promising avenue to increase mobility and improve many areas of their life. Policy decisions about the manner in which CAVs will be introduced onto public roads should be inclusive, and also rest on solid evidence regarding the needs of visually impaired individuals (Brinkley, Posadas, et al., 2019).

With the here presented study, we further understanding of CAV acceptance with quantitative evidence, giving voice to the expectations of citizens with visual impairments. As individuals with visual impairments have been found to be uniquely affected by the introduction of CAVs as a transportation option, it is of paramount importance to showcase which consequences they expect for their lives and society, and how this affects their evaluation of CAVs, and how this differs from those of sighted individuals.

Employing confirmatory factor analysis, we confirmed the factor structure previously found in investigation of CAV consequences, including safety, sustainability, efficiency, and privacy (Kacperski et al., 2021), and confirmed the inclusion of the additional factor independence, which was added due to its ubiquitous appearance in the literature of visually impaired citizens' insights about CAV introduction (Bennett et al., 2020; Brewer & Kameswaran, 2018; Brinkley et al., 2017).

In line with previous literature (Brinkley et al., 2017; Kempapidis et al., 2020), participants with visual impairments were significantly more optimistic about CAV introduction consequences and evaluations than panel participants. They on average expected CAVs to worsen privacy, but to improve sustainability, safety, efficiency and, crucially, independence.

Panel participants, as a control group, showcased a similar pattern except with neutral expectations regarding efficiency and independence.

Mediation analyses showed that sample differences in expected consequences for safety predicted both general and affective evaluations; differences in consequences for sustainability predicted affective evaluations and differences in the newly introduced factor independence predicted general evaluations. In summary, compared to a wider population, for citizens with visual impairments, safety and independence were the two mediators that better explained general evaluation, while sustainability and safety better explained affective evaluation.

Considerations of safety are in line with previous findings regarding CAV trust research in general (Choi & Ji, 2015; Kaur & Rampersad, 2018; Zhang et al., 2019), while increases in independence is a factor not commonly found in the general CAV acceptance literature. These two topics have been found to be of particular importance for people with visual impairments, due to higher needs for situational awareness, mobility assistance, and currently existing mobility barriers (Brewer & Kameswaran, 2018; Brinkley et al., 2017; Brito et al., 2018), with themes of improved quality of live frequently discussed due to current reliance on outsiders and a lack of flexible transportation options (Bennett et al., 2020; Brewer & Kameswaran, 2018). For individuals with visual impairments, there is often a tension between the needs for independence and for control and trust, and studies suggest that visually impaired individuals still desire support from real people (Fink, Alsamsam, et al., 2023). Therefore, in addition to purely technical solutions, the human factor will play a role in increasing acceptance through cooperation and emotional support.

And while CAV acceptance literature has sometimes included sustainability as a central topic (Acheampong et al., 2019; Nastjuk et al., 2020; Wu et al., 2019), it does not so far appear commonly as a theme in research with people with visual impairments outside of two instances regarding emissions and fuel efficiency (Brinkley & Gilbert, 2018; Miller et al., 2022). Further research in this direction could uncover interesting relationships between the goals to meet accessibility needs and environmental sustainability targets, as has been previously proposed in research on sustainable city and public transport planning (Capasso Da Silva et al., 2020; Corazza & Favaretto, 2019).

Finally, future research specifically in HMI and CAV design aspects should more closely examine which CAV features dictate the differing acceptance levels between sighted individuals and those who are visually impaired. As Fink et al. (2023) have shown, the needs regarding CAV equipment differ significantly; for example, visually impaired individuals place more importance on the availability of assistance with entering and exiting the vehicle.

**Policy recommendations**

CAV introduction will be majorly shaped by policymakers, industry partners and service providers, for example through implementations of such programs as the Inclusive Design Challenge (Inclusive Design, 2022) or EU funded inclusivity projects (European Comission, 2018). For policy and programs to be impactful, it is important to better understand personal and larger-scale perspectives of citizens with visual impairments, and their expectations about how CAVs will impact their lives and the world around them. But despite the high potential for inclusion for individuals with visual impairments, only a small percentage of published research is dedicated to their perceptions and expectations of autonomous mobility (Cavoli et al., 2017; Dicianno et al., 2021). Additionally, most studies are qualitative in nature. This provides a good basis for exploration and understanding of the perspectives of visually impaired citizens;

however, the lack of quantitative and experimental research might make it difficult to provide policy makers with the causal evidence necessary to enable essential regulatory changes.

Based on the results of the present study, which investigated personal and large-scale consequences of CAV introduction, there should be a particular effort to design safe, sustainable and accessible CAV solutions that enable inclusive transportation for all citizens, optimally employing the 7 principles of Universal Design (Mace, 1997). Advanced apps, booking systems, and vehicle equipment have previously been mentioned as hopeful options (Patel et al., 2021), but more importantly, the entire eco system and infrastructure should be designed with these three factors in mind.

Already, it has been pointed out that many avenues can be explored in the future to aid policymakers and other partners (Brinkley, Daily, et al., 2019; Fink et al., 2021). Overall, research should more thoroughly investigate acceptance related to various stages of the journey (Fink, Alsamsam, et al., 2023; Fink, Doore, et al., 2023), with both personal and larger-scale impacts taken into consideration and investigated (Dicianno et al., 2021), so that results from qualitative studies, participatory action design, and particularly experimental studies can enable a better evaluation of the consequences of CAV technology.

Emerging studies are expanding the research purview to pre-journey and post-journey experiences, acknowledging the challenges visually impaired individuals face when CAVs transport them to unfamiliar locales: the comprehension of these experiences is crucial for designing CAV systems that can adequately support their users before and after the actual travel, with particular attention to technologies that can provide situational awareness, control, and security (Fink et al., 2021; Fink, Alsamsam, et al., 2023).

Finally, given that ridesharing services are likely to be a key aspect of CAVs for the visually impaired population (Narayanan et al., 2020), not only the many in-vehicle elements will be critical, but the acceptance of pre-journey applications must be specifically explored (Brewer & Kameswaran, 2018; Fink et al., 2021; Kameswaran et al., 2018).

**Limitations and further research**

Due to their online and visual nature, quantitative survey-based online studies create an additional barrier for participation for individuals with visual impairments: a convenience sample procedure was employed to gather a sufficient number of responses, possibly fueling a self-selection bias. This might also be the reason for the higher educational attainment and income statistics we report in the participant section for individuals with visual impairments – it is likely that our participants are a subsample of the visually impaired population who have the means (in terms of time, technical equipment and technical knowledge) to answer an online survey. Results should therefore be generalized with care.

Due to a lack of statistical data available regarding age and gender of citizens with visual impairments in Germany, weighing of the sample was not possible (DBSV, 2021); we instead employed control variables. Analyses were restricted to working-age participants due to lack of access to participants of higher ages, who are very commonly diagnosed with visual impairments. While this reduces the generalizability beyond the working age population, we also believe that the results are very useful to draw conclusions about CAV acceptance for the presented population.

Additionally, analyses on subsamples related to sight loss (such as only blind or only partially sighted individuals) were not feasible due to the sample size. We relied on a self-selected assignment to these categories – it is possible that some participants had fully correctable visual impairments (e.g., with glasses). This means results should be interpreted as a broad recommendation across these categories. Unfortunately, recruitment in countries heavily affected by COVID-19 were cut short – so future studies should aim to include a wider range of locations to study between country or cultural differences. Further insights could be gained from such analyses into the needs of people with various visual impairments and across various cultures, helping to make CAV introduction more accessible and inclusive.

Mediation analyses were conducted with the expectation to deduce relative importance of the different factors for participants with visual impairments; the order in the mediation analysis – i.e. that expected consequences would predict evaluations – was theoretically founded based on findings in literature on CAV acceptance among individuals with visual impairments (Bennett et al., 2020; Brewer & Kameswaran, 2018; Brinkley et al., 2017, 2020; Brinkley & Gilbert, 2018; Miller et al., 2022). However, we cannot exclude the possibility that a latent positivity factor affects both, consequence factors and evaluation of CAVs. Future studies should attempt to disentangle the causal relationship further.

Modality is also an important factor to consider. In our study, we only asked participants about expected consequences from the role of a CAV user. Researchers should also explore the perspective of visual impairments as pedestrians or other road co-users.

**Conclusions**

In summary, we investigated the expected consequences of wide-spread CAV introduction by surveying the particular concerns of individuals with visual impairments and comparing them against expected consequences from a sample of panel participants. We found that an overall optimism is prevalent for citizens with visual impairments, both towards evaluations as well as all expected consequences of CAV introduction. Beside safety and sustainability as important factors in CAV acceptance, crucially, independence was found to have a sizeable effect, explaining why individuals with visual impairments show better attitudes towards CAVs. Our findings underscore the transformative potential of CAVs in enhancing the quality of life for individuals with visual impairments, offering new avenues for independence via mobility options. Insights gained can inform policymakers and CAV developers, ensuring that future advancements in autonomous vehicle technology are inclusive and cater to the diverse needs of all community members.